\documentclass{nature}
\usepackage[a4paper,top=2cm,bottom=2cm,left=2.cm,right=2.cm]{geometry}
\usepackage{graphicx}

\usepackage[bf, footnotesize]{caption}
\usepackage{amsmath}
\usepackage{amssymb}
\usepackage{hyperref}
\usepackage{txfonts}

\newcommand*\farcs{\ensuremath{\overset{\prime\prime}{.}}}
\newcommand*\fdg{\ensuremath{\overset{\circ}{.}}}

\newcommand{\citep}{\cite}

\spacing{1}

\begin{document}
\title{Closing the feedback-feeding loop of the radio galaxy 3C 84}

\title{Closing the feedback-feeding loop of the radio galaxy 3C 84}

\author{Tom Oosterloo$^{1,2}$, Raffaella Morganti$^{1,2}$ \& Suma Murthy$^{3}$}

\maketitle

\begin{affiliations}
 \item ASTRON, the Netherlands Institute for Radio Astronomy, Oude Hoogeveensedijk 4, 7991 PD Dwingeloo, The Netherlands.
 \item Kapteyn Astronomical Institute, University of Groningen, Postbus 800, 9700 AV Groningen, The Netherlands
 \item Joint Institute for VLBI ERIC, Oude Hoogeveensedijk 4, 7991 PD Dwingeloo, The Netherlands.

\end{affiliations}

\begin{abstract} 
Gas accretion by a galaxy's central super massive black hole (SMBH) and the resultant energetic feedback by the accreting active galactic nucleus (AGN) on the gas in and around a galaxy,  are two tightly intertwined but competing processes that  play a crucial role in the evolution of galaxies.  Observations of galaxy clusters have shown how the plasma jets emitted by the AGN heat the intra-cluster medium (ICM), preventing cooling of the cluster gas and thereby the infall of this gas onto the central galaxy.  On the other hand, outflows of multi-phase gas, driven by the jets, can cool as they rise into the ICM, leading to filaments of colder gas.  The fate of this cold gas is unclear, but it has been suggested it plays a role in feeding the central SMBH.  We present the results of re-processed CO(2-1) ALMA observations of the cold molecular gas in the central regions of NGC 1275, the central galaxy of the Perseus cluster and hosting the radio-loud AGN 3C 84 (Perseus A).  These data show, for the first time, in detail how kpc-sized cold gas filaments resulting from jet-induced cooling of cluster gas are flowing towards the galaxy centre and how they feed the circumnuclear accretion disc (100 pc diameter) of the SMBH. Thus, cooled gas can, in this way, play a role in  feeding  the AGN.  These results complete our view of the feedback loop of how an AGN can impact on its surroundings and how the effects from this impact maintain the AGN activity.   \label{Abstract} 
\end{abstract} 

\vskip0.3cm

Radio galaxies located in the centres of clusters are good candidates to explore the intricate connection between AGN feeding and feedback mechanisms and the role of these in galaxy evolution, thanks to the fact that they typically reside in environments rich in hot, warm, and cold gas\cite{Russell19}.  A pivotal object in this respect is NGC 1275, the brightest galaxy of the Perseus cluster and which hosts the radio loud AGN 3C~84 (Perseus A).  It is well known for the cavities and ripples in the hot cluster gas\cite{Fabian11} created by the jets of relativistic plasma ejected by 3C~84\cite{Pedlar90}, as well for its extensive web of optical emission-line filaments\cite{Minkowski57,Conselice01,Fabian08,Gendron18} which are also detected in warm\cite{Hatch05,Lim12} and cold molecular gas\cite{Salome06,Lim08,Ho09} as well as in colder X-ray gas\cite{Fabian11a}. In the centre of NGC 1275, a circum-nuclear disc (CND) with a radius of about 50 pc of molecular and ionised gas has been found\cite{Wilman05,Scharwachter13,Nagai19} with an orientation such that it is perpendicular to the inner radio jets. The kinematics of the outer parts of this CND suggests that the gas in this region is unsettled, possibly due to gas accreting onto the CND\cite{Scharwachter13}.  Large amounts of cold and warm molecular gas\cite{Hatch05,Salome06,Lim08,Ho09,Salome08,Salome11} has been detected in NGC 1275, out to radii as large as 50 kpc.  Most of the molecular gas is in a structure of about $20 \times 4$ kpc in size and oriented E-W, closely corresponding to that of the inner H$\alpha$\ emitting gas. In addition, the bulk of the molecular gas coincides with the locations of relatively cool X-ray gas lying between the X-ray cavities created by the radio jets\cite{Lim08}.

The intricate morphological connection between the radio lobes, the X-ray cavities, and the warm and cold gas suggests a close link between the processes that shape them.  Although here is still some debate about the details of how energy from the AGN is coupled into the ICM, the many studies on this object have resulted in a reasonable general consensus on what is happening in NGC 1275. Very briefly, the overall model is that the AGN in NGC 1275 emits jets of relativistic plasma to the north and to the south\cite{Pedlar90} which, while travelling through the hot cluster gas, create cavities and sound waves in the hot cluster gas\cite{Fabian11,Fabian17}, dumping energy int o the cluster gas in the process. This energy quenches, to some extent, the cooling flow of the hot ICM of the Perseus cluster which otherwise would occur\cite{Churazov02}. On the other hand, jet-driven outflows of multi-phase gas can cool as they rise, l eading to filaments of colder gas\cite{Hatch06,Qiu20} observed in H$\alpha$, molecular gas and in X-rays.  The overall layout of the various components of the gas in NGC 1275 has led, for example, Wilman et al.\cite{Wilman05}, Salom\'{e} et al.\cite{Salome06}\ and Lim et al.\cite{Lim08}\ to propose that the ionised and molecular gas fuels the CND which, in turn, is likely to be responsible for fuelling the AGN. However, while Lim et al.\cite{Lim08} observed filaments of molecular gas in the inner regions and Nagai et al.\cite{Nagai19} observed a central CND of cold molecular gas, no observations were yet able to simultaneously trace both these inner structures and explore, with high enough spatial resolution, whether any connection between them exists.  Here we present high-resolution ALMA observations that clearly show that kpc-sized filaments of molecular gas are indeed accreting onto the CND.

\section*{Molecular gas feeding 3C 84}

Past observations with a resolution of about 1 kpc of the cold molecular gas in the central regions of NGC 1275 have shown that the molecular gas is concentrated primarily in three radial filaments aligned in the east-west direction\cite{Lim08,Ho09}. The kinematics of some of these filaments could be consistent with radially infalling gas and these authors suggested based on this that the filaments are feeding a CND. However, the relatively low spatial resolution of their data do not allow to separate these large filaments from the CND and to establish whether the filaments are actually accreting onto the CND.

Using the much better resolution of ALMA ($\sim$50 pc), Nagai et al.\cite{Nagai19}\ were able to identify a rotating CND of cold molecular gas in NGC 1275.  However, due to bandpass calibration issues (see Methods), the data cube published by Nagai et al.\cite{Nagai19} is of relatively poor quality, suffering from strong imaging artefacts which prevent recovering the emission associated with the filaments detected by Lim et al.\cite{Lim08}\ and Ho et al.\cite{Ho09}\ and was thus not suitable to investigate the possible connection between the filaments and the CND.

Here we present results obtained after re-processing these ALMA observations, using a different bandpass calibration scheme (see Methods) which resulted in eliminating the imaging artefacts and thus in much improved image quality and much lower noise levels (see Fig.\ \ref{fig:ComparisonOld} and Methods). This has allowed us to clearly detect, with 50 pc resolution, the detailed distribution and kinematics of the molecular gas in the central kpc region.

\begin{figure*}
\centering
\includegraphics[width=\hsize,angle=0]{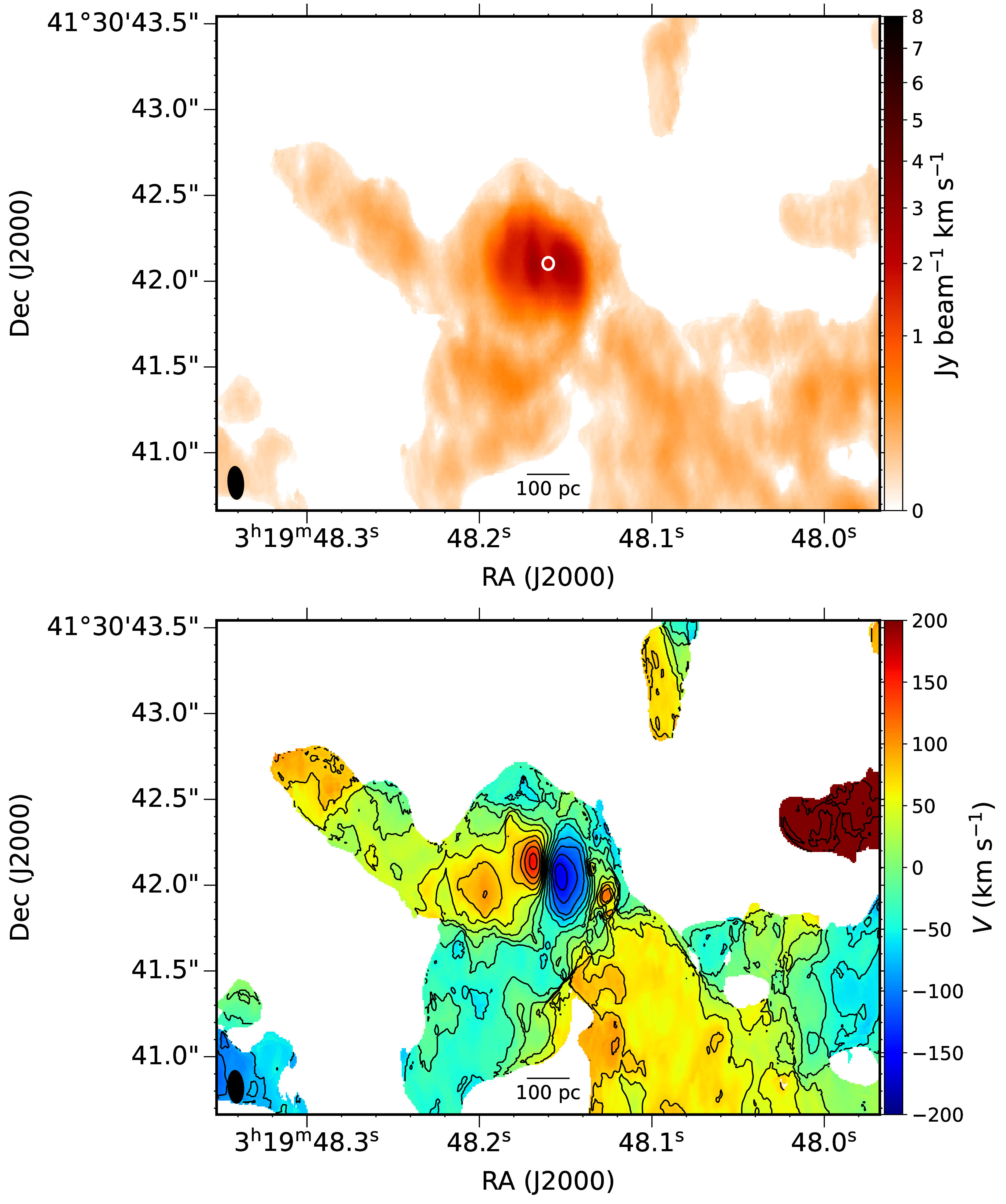}
\caption{{\bf Zoom-in of the CO emission and velocity field in the central regions of NGC~1275.} {\sl Top:} total intensity of the CO(2-1)\ of the CND and its immediate surroundings. The white circle marks the location of 3C~84.  {\sl Bottom:} velocity field of the same area showing the rotating CND and the kinematics of the filaments connecting to it. Contour levels are --200, --150, $\ldots$, 200 km~s$^{-1}$. The beam size is indicated in the bottom-left corner. }
\label{fig:velfieCND}
\end{figure*}

The main results are shown in Figs \ref{fig:TotInt} and \ref{fig:velfieCND} where we present the integrated line emission and the velocity field (see Methods) of the CO(2-1). These figures show that the so-called Inner and Western filaments as detected earlier by Lim et al.\cite{Lim08}\ and Ho et al.\cite{Ho09} split into several thinner filaments with different kinematics, thus revealing that the distribution and the kinematics of the molecular gas are much more complex than perceived from their data. Furthermore, they allow us to separate the filaments and the CND, something not possible from previous images.  Most importantly, the improvement of the new images allows us to study the connection between the CND and the molecular gas in the central regions of NGC 1275 and how the large filaments are feeding the CND.

The fact that the large filaments are all pointing towards 3C~84 suggests that they may be flowing towards or away from the CND. A number of properties support the idea that the gas must be infalling. For example, the fact that the gas immediately surrounding the CND is distributed mostly E-W with an overall organised kinematics, something that would not be expected if the gas is pushed outwards by the jets which are oriented N-S. The prevalence of infall is also indicated by the kinematics of the CND. In the centre, there is a clear kinematic signature of a fast rotating disc with position angle 68$^\circ$ and a diameter of $\sim$0\farcs3 ($\sim$100 pc) as previously reported by Nagai et al.\cite{Nagai19} and detected earlier in warm H$_2$\cite{Wilman05,Scharwachter13} using shock-excited ro-vibrational H$_2$ emission lines in the near infrared. This was interpreted by Scharw{\"a}chter et al.\cite{Scharwachter13} as the outer parts of a collisionally excited turbulent accretion disc.  However, the asymmetry of the velocity field of the CND at somewhat larger radii shows that its inner fast-rotating region is embedded in a slightly larger disc-like structure with much less regular kinematics indicating accretion of gas onto the CND.  In fact, as was observed for the warm H$_2$\cite{Scharwachter13}, on the eastern side of the CND, at a radius of about 0\farcs5, the kinematic major axis of the velocity field of the cold molecular gas shows a twist from 68$^\circ$ towards larger posit ion angles (see zoom-in of the central region in Fig.\ \ref{fig:velfieCND}).  Similarly, about 1 arcsec SW of the CND, counter-rotating velocities with respect to the W side of the CND are seen. Similar velocities are also detected in warm H$_2$\ in this region\cite{Scharwachter13}. This was interpreted by Scharw{\"a}chter et al.\cite{Scharwachter13} as being due to a gas streamer accreting onto the CND. Our data now show that this gas seen in warm H$_2$\ is only the tip of a much larger filament (fila ment C in Fig.\ \ref{fig:TotInt}). A third stream of gas (D in Fig.\ \ref{fig:TotInt}) at velocities closer to systemic is seen connecting to the CND from the south-east.

\begin{figure}
\centering
\includegraphics[width=\hsize]{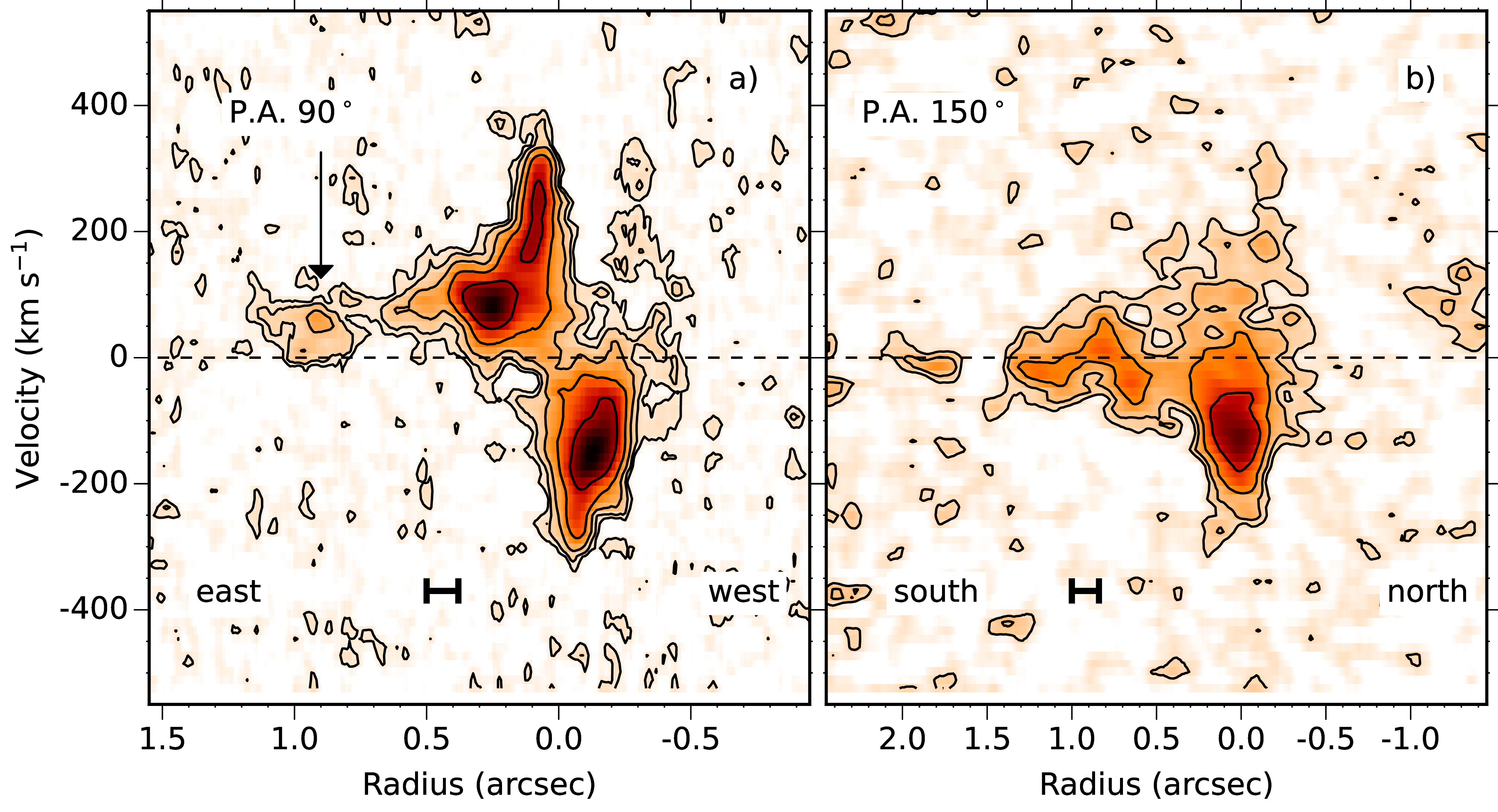}
\caption{{\bf Position-velocity plots taken along   CO(2-1)\ filaments accreting onto the CND.}
 {\bf a)} Position-velocity plot through the centre along position angle 90$^\circ$. The bright, inner part corresponds to the CND which was earlier studied by Nagai et al.\cite{Nagai19}. On the eastern side of the CND, the velocities drop very quickly towards the systemic velocity and this is what causes the twist in the velocity field of the CND. The arrow indicates a filament that is connected to the outer CND and is likely responsible for the disturbed kinematics of the outer parts of the CND. 
{\bf b)} Position-velocity plot along position angle 150$^\circ$ and centred on the W part of the CND, showing how filament D is connected to the CND also in velocity. Velocities are relative to the systemic velocity of NGC 1275 indicated by the dashed line. Contour levels are 1.5, 3, 4.5, 6, ...  $\sigma$. The scale bar indicates the spatial resolution.}
\label{fig:slicesCND}
\end{figure}

The connection between some of the filaments and the CND is further illustrated by the kinematics of the molecular gas along these filaments down to the CND. We show in Fig.\ \ref{fig:slicesCND} position-velocity plots along two streams that seem to feed the CND. The left panel shows the kinematics of the CND along position angle 90$^\circ$. The inner bright emission traces the kinematics of the CND which has been discussed in detail in Nagai et al.\ (2019)\cite{Nagai19}. It is clear that the distribution and kinematics of the bright inner disk is very asymmetric, indicating that this inner disk is disturbed. It can also be seen that on the eastern side of the brighter outer disk the velocities drop very quickly towards the systemic velocity, resulting in the twist in the velocity field mentioned above. But the figure also shows that this outer CND is connected to a filament further to the east, suggesting this filament is flowing onto the CND and is responsible for the disturbed kinematics. The right panel of Fig.\ \ref{fig:slicesCND} shows the position-velocity diagram along filament D (position angle 150$^\circ$) centred on a position 0\farcs3 W of the centre, clearly showing that this filament is connected in velocity to the W side of the CND, suggesting a physical connection between the two. Finally, the much larger mass of the filaments compared to the CND further (indirectly) supports the infall of gas onto the CND.

\begin{figure}
\centering
\includegraphics[width=\hsize,angle=0]{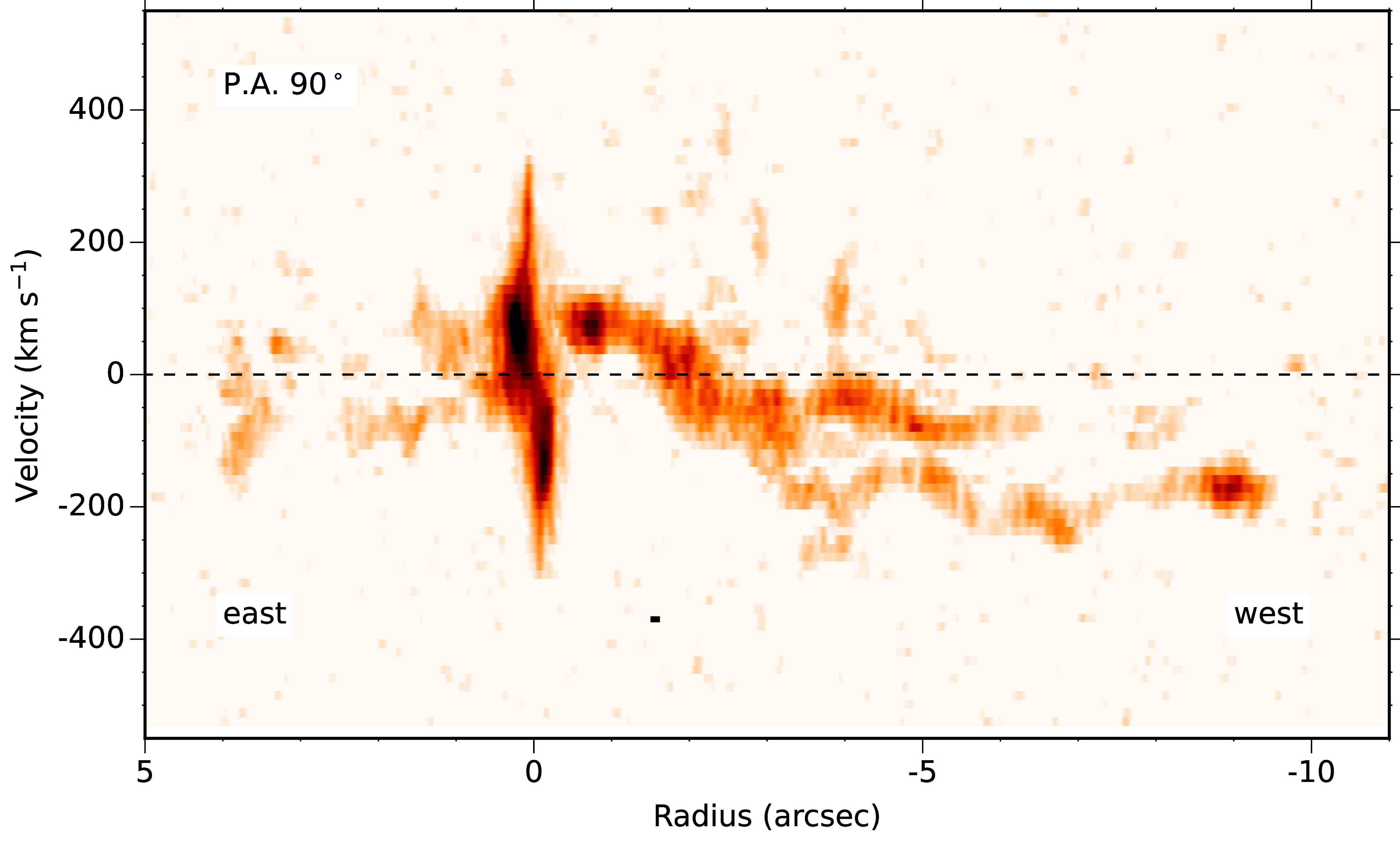}
\caption{ {\bf Position-velocity plot of the CO(2-1)\ emission after integrating the data cube in declination}. The structure with a very wide range of velocities corresponds to the CND. On both the eastern and the western side negative as well as positive velocities are seen, indicating there is no overall rotational pattern. Velocities are relative to the systemic velocity of NGC 1275.  }
\label{fig:sliceInt}
\end{figure}

Figure \ref{fig:TotInt} shows that on larger scales, the kinematics of the molecular gas is quite complicated and that there is a large range in the velocities, even locally, of the different filaments and clouds. This is further illustrated in Fig.\ \ref{fig:sliceInt} which shows the position-velocity plot along position angle 90$^\circ$ after integrating the data cube in declination.  This figure shows that on both sides gas with positive and negative velocities is found. As was observed at lower resolution for the molecular gas on even larger scales\cite{Salome06} and for the H$\alpha$\ filaments\cite{Gendron18}, no overall rotation is present while it also shows that the kinematics is more complicated than pure radial infall. Thus, individual filaments appear to have some angular momentum and may follow a more conical helix trajectory while falling towards the centre, as in chaotic cold accretion models\cite{Gaspari17}.  However, the accretion itself is not chaotic but organised in the form of filamentary channels along which cold gas is funnelled into the central regions.

The cold H$_2$\ masses of filaments A, B and C are $4.4 \pm 0.2\times 10^8$, $5.2 \pm 0.3 \times 10^8$, and $3.1 \pm 0.2 \times 10^8$ $M_\odot$\ respectively if a standard Galactic ISM conversion factor\cite{Bolatto13} is used, but are a factor 5 lower if a conversion factor perhaps more appropriate for conditions near an AGN is used (see Methods). Several smaller clouds with masses down to a few times $10^7$ $M_\odot$\ are also detected, at various velocities.  Although there is a large uncertainty in which CO-to-H$_2$\ conversion factor to use, we estimate that the disposition rate of cold gas onto the CND to be in the range 20--200 $M_\odot$ yr$^{-1}$\ (see Methods). This compares reasonably well with the disposition rate of at least 30 $M_\odot$ yr$^{-1}$\ of the X-ray cooling flow derived by Bregman et al.\cite{Bregman06}\ from O {\sc vi} emission lines for a 30$^{\prime\prime}$ central square region, or the estimated cooling rate in the H$\alpha$\ filaments of about 100 $M_\odot$ yr$^{-1}$\ found by Fabian et al.\cite{Fabian11}. The large filaments west of 3C 84 are very elongated. The longest filament (marked A in Fig.\ \ref{fig:TotInt}) has a length of about 2.7 kpc, but has a width of only $\sim$150 pc, while filament B is somewhat shorter and wider ($\sim$1.8 kpc long and 230 pc wide). The long, linear shape is reminiscent of that of the long, thin H$\alpha$\ filaments seen at other locations in NGC~1275\cite{Conselice01,Fabian08}, suggesting {they may be related to similar phenomena}.

\begin{figure}
\centering
\includegraphics[width=\hsize,angle=0]{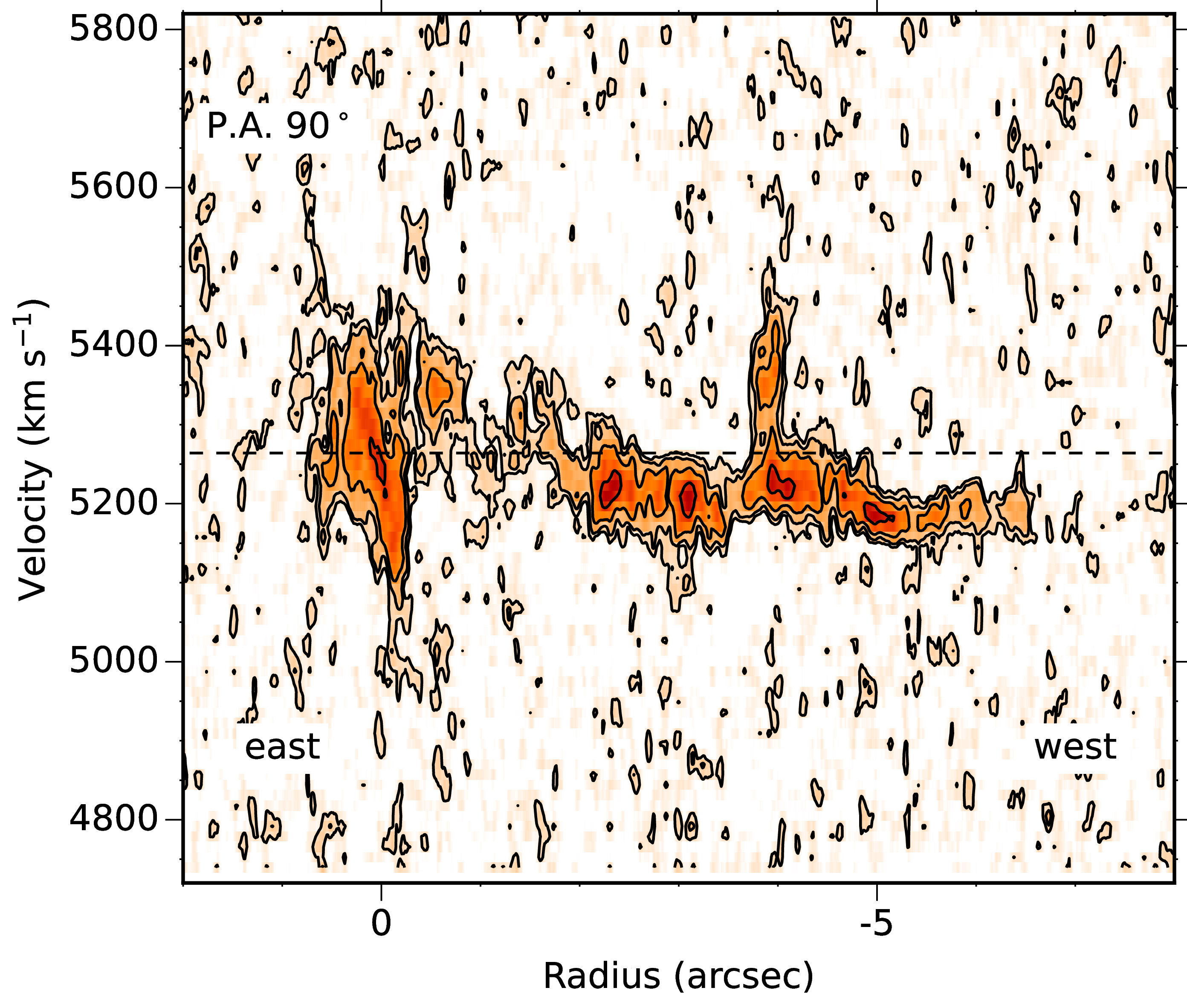}
\caption{ {\bf Position-velocity plot along filament B showing local anomalous velocities.} Position-velocity plot (position angle $90^\circ$) after integrating the data cube in declination over 0\farcs45 of filament B.  The dashed line indicates the systemic velocity. Contour levels are 1.5, 3, and 6 $\sigma$}
\label{fig:jump}
\end{figure}

Finally, some puzzling features are observed in the kinematics of some of the larger filaments. The velocity field shown in Fig.\ \ref{fig:TotInt} shows that at a few locations in the western filaments there are large, local discontinuities in the velocities.  This is illustrated in Fig.\ \ref{fig:jump} which presents a position-velocity diagram along filament B, showing the most prominent of such velocity jumps. This figure shows that the velocities of the filament vary smoothly with radius.  However at about 4$^{\prime\prime}$ from the centre, there is a small cloud which is displaced by about 200 km~s$^{-1}$\ from the main filament but seems to be connected to it in velocity. It is not clear what causes these abrupt velocity features. One possibility could be that they are the result from a gravitational interaction of a filament with a relatively massive object, such as one of the super star clusters which are known to be present in the region where we observe the filaments and which have sizes $\lesssim$20 pc and masses up to $10^7$ $M_\odot$\cite{Holtzman92,Lim22}.

\section*{Closing the feedback loop}

With the improved calibration of the ALMA observation, we were able to obtain two main results for better understanding the connection between the various structures observed in cold molecular gas and their role for the cycle of activity in 3C~84.  Firstly, the improved sensitivity has allowed us to trace -- in the region of about 4 kpc radius around 3C~84 -- the detailed morphology and kinematics of the filaments of cold molecular gas and and how they connect to the CND. The high spatial resolution of the ALMA observations (50 pc) show that the structure of the filaments is much more complicated than seen in earlier, low-resolution data\cite{Lim08,Ho09}. Secondly, the data clearly show how kpc-sized filaments flow towards the CND and we see direct signatures of how filaments of cold gas are accreting onto the CND, building it up in mass.  Other large filaments at larger radii will do so in the near future. This on-going formation of the CND is further confirmed by the non-regular kinematics of the outer regions of the CND which suggest that the gas there is not in a kinetically relaxed state and, given the arrival on new filaments, will likely never reach a fully relaxed condition.

The unambiguous evidence of the connection between filaments and CND allows us to complete the picture of the gas cycle in Perseus A sketched in the introduction based on data in many wavebands taken from the literature.  Several authors have suggested that in NGC 1275 we observe a mixture of heating and cooling processes, both driven by the impact of the AGN on the gaseous medium surrounding it, and how these processes could be connected to the fuelling of the AGN\cite{Salome06,Lim08,Scharwachter13}, much in the way as predicted by models of AGN-driven heating and condensation models\cite{Gaspari17}. The observed spatial correlation between the large-scale distribution of the molecular and ionised gas with that of the colder X-ray gas, as well as the anti-correlation of these tracers with the radio lobes, suggests that multi-phase outflows, driven by the jets, cools as they rise, leading to formation of colder gas filaments\cite{Qiu20}.  Earlier studies have proposed that these filaments may ultimately fall back and accrete onto the CND and eventually fuel the AGN\cite{Salome06,Lim08}, similar as in models of AGN-driven accretion models\cite{Gaspari17}. This would close the feedback and fuelling loop of the AGN.  Our results from ALMA provide the observational evidence that this process is indeed on-going.

Our findings can also have interesting implications for the radio jet.  The unsettled kinematics of the CND suggest that it is likely that, over time, the orientation of the CND changes because the orientation of the angular momentum of the gas filaments accreting onto the CND can be different from that of the CND. Assuming that the inner radio jet stays always perpendicular to the inner regions of the CND, this means that the jet orientation will also change over time. Indications, from both large\cite{Pedlar90} and small\cite{Nagai10} scales, have been found that this indeed has happened in the past. For example, the direction of movement (on pc scales) of the new jet component associated with the most recent (2005) outburst differs from that of the pre-existing component by $\sim 40^\circ$\cite{Nagai10}.  It has been suggested that some of these changes in orientation (in particular on the pc scale) are due to the jets interacting with clouds in the ISM of NGC~1275\cite{Kino21}. However, if our interpretation of our observations is correct, these changes might instead be due to the CND changing orientation over time. On the other hand, much larger changes in orientation from an approximately N-S direction are not likely as accretion onto the CND is primarily in the E-W direction.

The large filaments we detect, and which will be forming the CND in the near future, will fall onto the CND mostly from the east-west direction. In addition, the main structure of ionised and molecular gas on the scale of several kpc is also oriented east-west\cite{Conselice01,Salome06}. Therefore, when in the more distant future, gas from even larger radii will be deposited onto the CND, it will also do this mostly in the east-west direction. This means that the CND, and thus the jet, may change orientation, but the jets will still be launched largely in the north-south directions, meaning that the orientation of the radio structure will remain more or less north-south. Furthermore, the ALMA results suggest that the continued accretion of cold molecular gas from the filaments onto the AGN ensures triggering of new jet components as recently observed\cite{Nagai10}, which will maintain the radio jet active as known to be the case in central radio galaxies of cool-core clusters\cite{Hogan15} like 3C~84. Finally, the continuous but not uniform flow of gas towards the CND can also be relevant in explaining the strong long-term variability of 3C~84\cite{ODea84,Dutson14,Paraschos23}.

\clearpage
\begin{addendum}

 \item[Data availability] The calibrated data cube is available from\\ 
\href{https://astrodrive.astro.rug.nl/index.php/s/g1y6QlCeGFd5XiG}
{https://astrodrive.astro.rug.nl/index.php/s/g1y6QlCeGFd5XiG}, 
or from the corresponding author on reasonable request. 
  
\item[Code availability] The data were reduced using the publicly available 
software Miriad\cite{Sault95} and Sofia-2\cite{Westmeier21}.

\item This work is based on re-processing of the ALMA observations carried out under project number  2017.0.01257.S and  which were published in original form by Nagai et al.\ \cite{Nagai19}
 
\item[Author contributions] TO and RM conceived the project. TO reduced the data. 
TO, RM and SM carried out the analysis and wrote the manuscript. 
All the authors discussed the results and commented on the manuscript.
  
\item[Competing Interests] The authors declare that they have no competing  interests.

\item[Correspondence]Correspondence and requests for materials should be 
addressed to Tom Oosterloo~(email: oosterloo@astron.nl).

\end{addendum}

\clearpage

\begin{figure} \centering
  \includegraphics[width=\hsize]{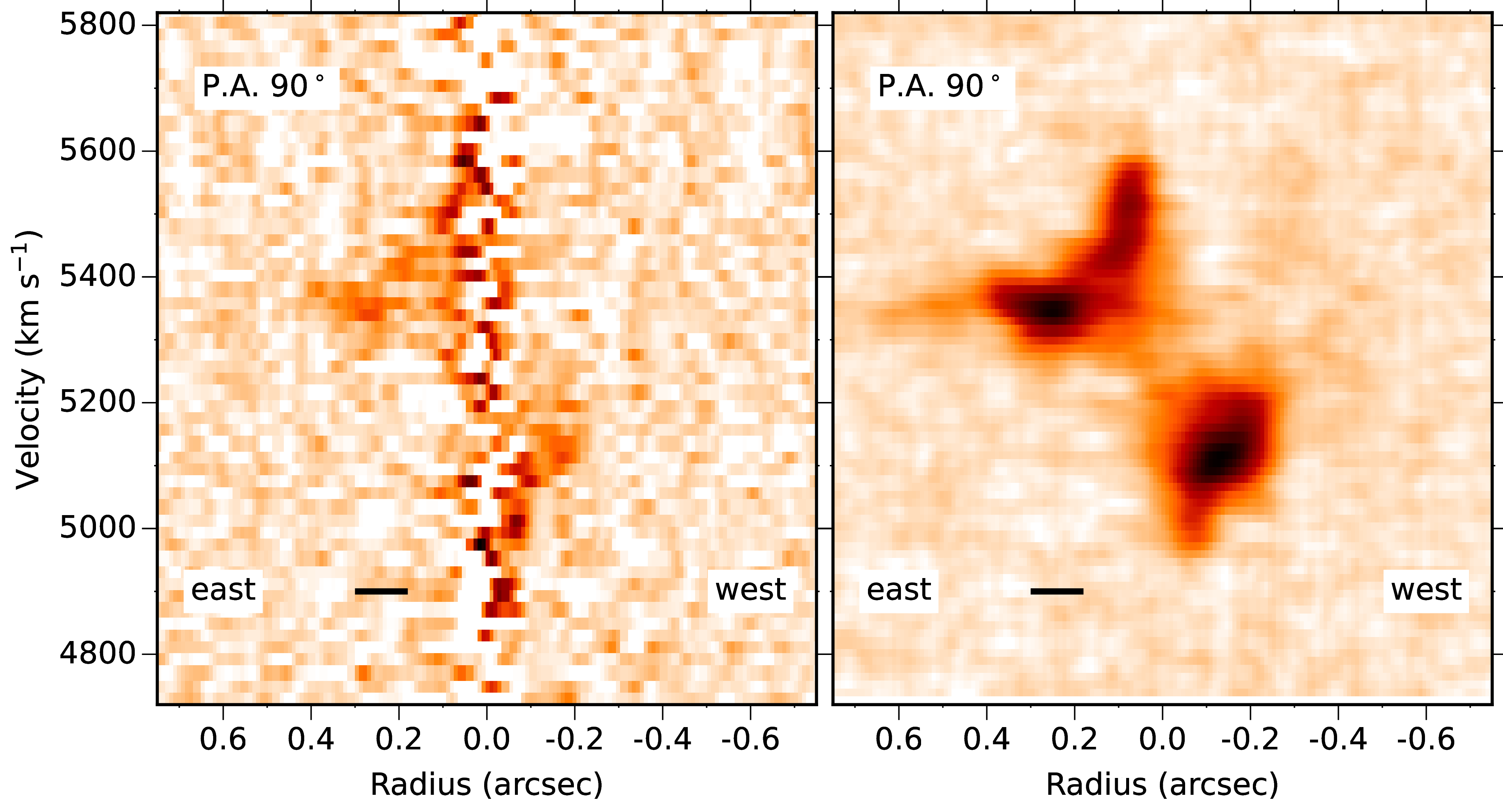}
  \caption{{{\bf Position-velocity plots illustrating the improvement of the data after re-processing}. 
  \sl Top:} position-velocity map taken along position angle $90^\circ$ centred at the position of 3C~84 taken from the archival data. {\sl Bottom:} Same as on the left, but now using the re-processed data presented in this paper. The intensity scaling for the two panels is the same. }
\label{fig:ComparisonOld}
\end{figure}

\begin{methods}

\subsection{Target information.} 

NGC 1275 is the brightest galaxy of the Perseus cluster and hosts the radio-loud AGN 3C~84 (Perseus A). In this paper we assume a redshift for NGC 1275 of $z = 0.01756; V_{\rm hel} = 5264$ km~s$^{-1}$. Assuming a Hubble constant of $H_\circ = 69.6$ km~s$^{-1}$\ Mpc$ ^{-1}$, $\Omega_{\rm M} = 0.286$ and $\Omega_{\rm vac} = 0.714$, this gives a scale of 0\farcs1 = 36 pc and 1 kpc $\simeq$ $3^{\prime\prime}$, and a luminosity distance of 76.2 Mpc.

\subsection{Observations and data reduction.}

This paper is based on re-processing the data taken for ALMA project 2017.0.01257.S of which the original data were published by Nagai et al. \cite{Nagai19}. The quality of the data cube produced by the standard ALMA pipeline as published by Nagai et al.\cite{Nagai19} is strongly limited by the poor quality of the bandpass calibration, as was already noted by Nagai et al.\cite{Nagai19}. The main reason for this is that there are no calibrator sources on the sky that are significantly stronger than 3C~84 ($\sim$7 Jy), the strong radio source hosted by the target source NGC~1275.  Therefore, unless one is willing to spend more time observing a calibrator than the target, the spectral dynamic range of the data will be limited. For these observations, the flux density of the bandpass calibrator used (J0237+2848) is about 1 Jy while that of 3C~84 is roughly 7 Jy. In addition, the bandpass calibrator was observed for only 5 minutes, compared to about 1 hr for 3C~84.  These two issues mean that the S/N of the bandpass calibration is insufficient to obtain the required spectral dynamic range (see \cite{Nagai19} for a detailed discussion of this).  The impact of this is elevated noise in the data cube at the location of 3C~84. In addition, it also leads to strong, channel dependent continuum residuals covering the entire channel images because of the limited precision of the continuum subtraction, complicating the detection of faint and/or extended emission.

To improve the calibration of these data, we reprocessed the archival ALMA visibility data. All our data reduction was done using the MIRIAD software\cite{Sault95}, starting with the calibrated $uv$ data produced by the ALMA pipeline. Before attempting to improve the bandpass calibration, we applied a single phase-only self-calibration to the visibilities using a point model for 3C 84.

Our strategy to improve the spectral dynamic range of the data is based on using 3C 84 itself (being an ALMA calibrator)  as the bandpass calibrator. The complication of this approach is, of course, that  line emission present in the data will lead to errors in the resulting bandpass. These, in turn, give imaging errors of the line emission if this bandpass is used. However, the level of the continuum is much higher than that of the line emission and therefore the errors induced by this way of calibrating the data are small. We further limited the impact of the line emission on the bandpass by using an iterative process of removing the line emission from the $uv$ data, as well as only using long baselines for the bandpass calibration. We started by determining, in the standard way, a bandpass solution using the data on 3C~84 and selecting only the $uv$ data on baselines longer than 500 k$\lambda$ in order to reduce the impact of extended line emission. This limit means that for the bandpass calibration we only consider spatial scales smaller than $\sim$0\farcs4 (i.e.\ $\sim$2--3 times the resolution). The implication is that our final data cube is likely incorrect within $\sim$0\farcs2 from 3C 84 (i.e.\ the inner part of the CND), but that the data are correct outside this region; it turns out that most of the line emission we detect is on much larger scales.  Another implication of our calibration method is that any absorption signature that would possibly be present is calibrated away and that it is not possible to extract any information about absorption against the continuum of 3C 84.

Although the bandpass resulting from the first iteration is only approximate, it does give an improvement of the data cube. This allows us to subtract most of the line emission from the $uv$ data, making it possible to improve the bandpass calibration in a next iteration using the line-subtracted $uv$ data The process of modelling the line emission and subtracting it from the visibilities, followed by a new determination of the bandpass was done twice, after which successive iterations did not improve the quality of the data anymore. The final data cube we use in this paper was made using natural weighting and with channels of 13.1 km~s$^{-1}$\ wide to which Hanning smoothing was applied, giving a velocity resolution of 26.1 km~s$^{-1}$. The beam size is 0\farcs19 $\times$ 0\farcs10 (P.A.\ = 4\fdg8), corresponding to 68 $\times$ 36 pc, and the noise level in the channels is 0.36 mJy~beam$^{-1}$. This is an improvement of a factor 3 over that of the original data cube (1 mJy~beam$^{-1}$\ for a spectral resolution of 20 km~s$^{-1}$). This is due to the absence in the new cube of the large-scale continuum residuals that are present in the channels of the original data cube left after continuum subtraction.

To illustrate the improvement in data quality, we show in Fig.\ \ref{fig:ComparisonOld} position-velocity plots of the CND of both the archival data and of the re-processed data. We reiterate that our bandpass calibration procedure will calibrate away emission very close to the strong continuum emission of 3C 84. Therefore it is likely that the fact the new data shows little emission in the very centre is an artefact of our calibration procedure. However, comparison of the emission only a few beams away from 3C 84 shows very good correspondence between the two data sets, which suggests that the imaging errors due to our calibration procedure are limited to the inner 0\farcs1 or less and that the CO emission beyond this radius is imaged with good fidelity.

\subsection{Moment images}

The integrated line image and the velocity field were made using the Sofia-2\cite{Westmeier21} software. This software does a multi-resolution analysis of the data cube by smoothing the data cube spatially and in velocity using a set of kernels of different sizes, followed by an analysis of the noise at each scale in order to select regions of emission. We have used the Smooth+Clip finder as implemented in Sofia-2 to create a mask including all the emission to apply to the data cube for the calculations of the moments. This involves smoothing the data cube both spatially and in velocity by a set of kernels and for each level of smoothing a mask is made based on the noise level of the cube at that smoothing level. The final mask that is applied to the original data cube is obtained by merging these masks. The scales of the spatial Gaussian kernels we used are 0, 3, 6, 12 and 24 pixels and of the velocity boxcar kernels 0, 3, and 7 channels. The noise cutoff used was 4 $\sigma$.

\subsection{Molecular gas masses}

The flux integral over the entire field of view is $112.9 \pm 6.0$ Jy km~s$^{-1}$\ with the bulk of the emission coming from the central region and the large filaments west and east of 3C 84. The flux integral of the CND is $22.2 \pm 1$ Jy km~s$^{-1}$, of the three main filaments A, B and C as indicated in Fig.\ \ref{fig:TotInt} $17.6 \pm 0.8$, $20.7\pm 1.0$ and $12.6\pm 0.6$ Jy km~s$^{-1}$\ respectively. The largest recoverable scale of the ALMA array used in the observations is 1\farcs3. The width of most of the emission in a single channel is of this size or smaller which suggests that we have likely recovered most of the CO(2-1)\ emission.

A comparison with the result obtained by Ho et al.\cite{Ho09} is not straightforward, given the very different resolution of the two data sets. The combined flux integral of their inner and the western filaments, which correspond to most of the emission in our data, is $106.0 \pm 3.4 $ Jy km~s$^{-1}$.  The flux integral of their western filament is $34.1 \pm 2.3$ Jy km~s$^{-1}$, which roughly corresponds to our filaments A and B which have a combined flux integral of $38.3 \pm 1.9$ Jy km~s$^{-1}$. The primary beam of the SMA observations of Ho et al.\ is 55 arcsec while it is 25 arcsec of our ALMA observations. Therefore what is called the eastern filament by Ho et al.\ is located well beyond the half-power point of the ALMA primary beam and only some faint emission is detected at that location in our observations.

It is a priori not clear which CO--to--H$_2$\ conversion factor to use since the conditions in the molecular gas are likely quite different from that in the standard galaxy ISM\cite{Lim17}. Using $\alpha_{\rm CO} = 4.3\ M_\odot$/(K km s$^{-1}$ pc$^2$)\ and $S_{\rm CO(2-1)} = 2.4\ S_{\rm CO(1-0)}$, which are characteristic of a standard ISM\cite{Bolatto13}, this gives a total H$_2$\ mass of $2.79 \pm 0.14\times 10^9$ $M_\odot$, and of filaments A, B and C the H$_2$\ masses are $4.4 \pm 0.2\times 10^8$, $5.2 \pm 0.3 \times 10^8$, and $3.1 \pm 0.2 \times 10^8$ $M_\odot$\ respectively. The H$_2$\ mass of the CND is $5.6 \pm 0.3 \times 10^8$ $M_\odot$.  The smallest clouds in our data have a flux integral of $\sim$0.6 Jy km~s$^{-1}$, corresponding to a mass of $1.5 \times 10^7$ $M_\odot$. If one instead assumes $\alpha_{\rm CO} = 0.8\ M_\odot$/(K km s$^{-1}$ pc$^2$), which could more appropriate for conditions found near an AGN, all masses decrease by a factor 5.4.

\subsection{Accretion rates}

A rough estimate of the rate with which the cold molecular  gas is accreting onto the CND can be made using
\begin{equation}
\dot{M} = \frac{M_{\rm gas}}{t_{\rm dyn}} = M_{\rm gas} \cdot \left(\frac{r}{v}\right)^{-1}
\end{equation}
where $r$ is  the distance from the centre of NGC 1275 and $v$ the infall velocity.
The gas mass of the three filaments inside a region with a radius of 0\farcs7 ($\sim$250 pc)  which appear to be accreting onto the CND (thus excluding the bright inner region) is about $5 \times 10^8$ $M_\odot$\ for $\alpha_{\rm CO} = 4.3\ M_\odot$/(K km s$^{-1}$ pc$^2$). In the models  of Lim et al.\cite{Lim08} for the radial infall of the filaments towards the CND, they do so  with velocities of the order of 200 km~s$^{-1}$. If we assume such an infall velocity, this gives an estimated infall rate of about 200 $M_\odot$ yr$^{-1}$. Assuming $\alpha_{\rm CO} = 0.8\ M_\odot$/(K km s$^{-1}$ pc$^2$)\ this reduces to $\sim$40 $M_\odot$ yr$^{-1}$. Similarly, the combined mass of the two largest filaments (A and B), which are about 2 kpc from 3C 84 and may fall onto  the CND in the near future, is $9.5 \times 10^8$ $M_\odot$. Again assuming an infall velocity of 200 km~s$^{-1}$, this gives an infall rate of $\sim$90 $M_\odot$ yr$^{-1}$\ or $\sim$20 $M_\odot$ yr$^{-1}$, depending on the conversion factor assumed. Assuming the same CO-to-H$_2$ conversion factor for the filaments and the CND and the infall velocity of 200 km~s$^{-1}$\ used above, it would take  just several times $10^6$ yr to build the CND. 

\end{methods}

\clearpage

\begin{figure}
\centering
\includegraphics[width=\hsize,angle=0]{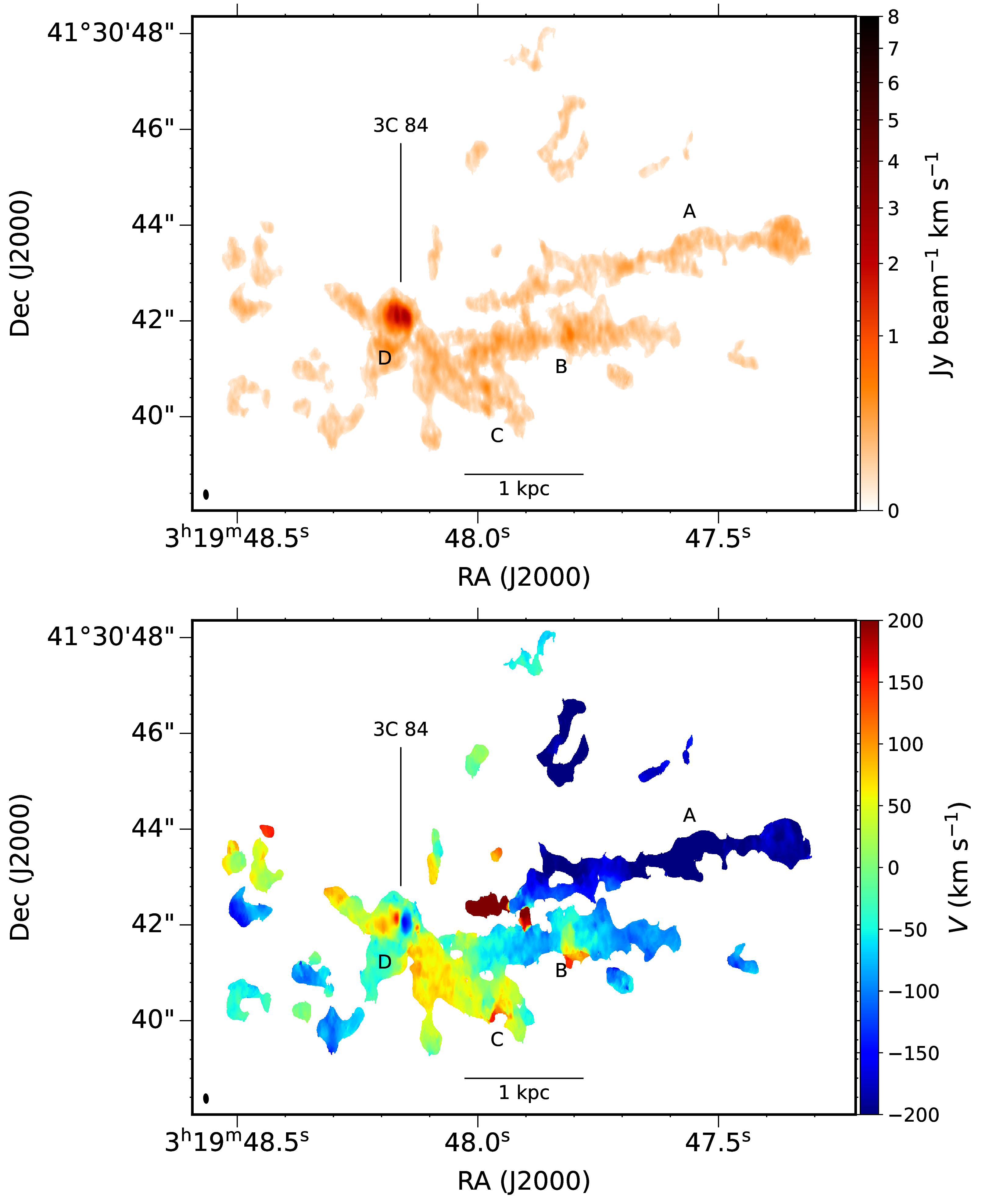}
\caption{{\bf Total intensity image and velocity field of the CO(2-1)\ in the central regions of NGC 1275}. {\sl Top)} Total intensity image of the CO(2-1)\ showing the CND in the centre of a complex network of filaments and clouds. {\sl Bottom)} Velocity field of the CO(2-1)\ over the same region as the top panel, showing the complicated velocity structure of the filaments and the clear kinematic signature of the inner part of the CND. Velocities are relative to the systemic velocity of 3C~84 The position of 3C 84 is indicated, as well as the main filaments discussed in the text.  The beam size is indicated in the bottom-left corner.}
\label{fig:TotInt}
\end{figure}

\clearpage

\begin{thebibliography}{10}
\expandafter\ifx\csname url\endcsname\relax
  \def\url#1{\texttt{#1}}\fi
\expandafter\ifx\csname urlprefix\endcsname\relax\def\urlprefix{URL }\fi
\providecommand{\bibinfo}[2]{#2}
\providecommand{\eprint}[2][]{\url{#2}}

\bibitem{Russell19}
\bibinfo{author}{{Russell}, H.~R.} \emph{et~al.}
\newblock \bibinfo{title}{{Driving massive molecular gas flows in central
  cluster galaxies with AGN feedback}}.
\newblock \emph{\bibinfo{journal}{Mon. Not. R. Astron. Soc.}} \textbf{\bibinfo{volume}{490}},
  \bibinfo{pages}{3025--3045} (\bibinfo{year}{2019}).
\newblock \eprint{astro-ph/1902.09227}.

\bibitem{Fabian11}
\bibinfo{author}{{Fabian}, A.~C.} \emph{et~al.}
\newblock \bibinfo{title}{{A wide Chandra view of the core of the Perseus
  cluster}}.
\newblock \emph{\bibinfo{journal}{Mon. Not. R. Astron. Soc.}} \textbf{\bibinfo{volume}{418}},
  \bibinfo{pages}{2154--2164} (\bibinfo{year}{2011}).
\newblock \eprint{astro-ph/1105.5025}.

\bibitem{Pedlar90}
\bibinfo{author}{{Pedlar}, A.} \emph{et~al.}
\newblock \bibinfo{title}{{The radio structure of NGC 1275.}}
\newblock \emph{\bibinfo{journal}{Mon. Not. R. Astron. Soc.}} \textbf{\bibinfo{volume}{246}},
  \bibinfo{pages}{477} (\bibinfo{year}{1990}).

\bibitem{Minkowski57}
\bibinfo{author}{{Minkowski}, R.}
\newblock \bibinfo{title}{{Optical investigations of radio sources
  (Introductory Lecture)}}.
\newblock In \bibinfo{editor}{{van de Hulst}, H.~C.} (ed.)
  \emph{\bibinfo{booktitle}{Radio astronomy}}, vol.~\bibinfo{volume}{4},
  \bibinfo{pages}{107} (\bibinfo{year}{1957}).

\bibitem{Conselice01}
\bibinfo{author}{{Conselice}, C.~J.}, \bibinfo{author}{{Gallagher}, I.,
  John~S.} \& \bibinfo{author}{{Wyse}, R. F.~G.}
\newblock \bibinfo{title}{{On the Nature of the NGC 1275 System}}.
\newblock \emph{\bibinfo{journal}{Astron. J.}} \textbf{\bibinfo{volume}{122}},
  \bibinfo{pages}{2281--2300} (\bibinfo{year}{2001}).
\newblock \eprint{astro-ph/0108019}.

\bibitem{Fabian08}
\bibinfo{author}{{Fabian}, A.~C.} \emph{et~al.}
\newblock \bibinfo{title}{{Magnetic support of the optical emission line
  filaments in NGC 1275}}.
\newblock \emph{\bibinfo{journal}{Nature}} \textbf{\bibinfo{volume}{454}},
  \bibinfo{pages}{968--970} (\bibinfo{year}{2008}).
\newblock \eprint{astro-ph/0808.2712}.

\bibitem{Gendron18}
\bibinfo{author}{{Gendron-Marsolais}, M.} \emph{et~al.}
\newblock \bibinfo{title}{{Revealing the velocity structure of the filamentary
  nebula in NGC 1275 in its entirety}}.
\newblock \emph{\bibinfo{journal}{Mon. Not. R. Astron. Soc.}} \textbf{\bibinfo{volume}{479}},
  \bibinfo{pages}{L28--L33} (\bibinfo{year}{2018}).
\newblock \eprint{astro-ph/1802.00031}.

\bibitem{Hatch05}
\bibinfo{author}{{Hatch}, N.~A.}, \bibinfo{author}{{Crawford}, C.~S.},
  \bibinfo{author}{{Fabian}, A.~C.} \& \bibinfo{author}{{Johnstone}, R.~M.}
\newblock \bibinfo{title}{{Detections of molecular hydrogen in the outer
  filaments of NGC1275}}.
\newblock \emph{\bibinfo{journal}{Mon. Not. R. Astron. Soc.}} \textbf{\bibinfo{volume}{358}},
  \bibinfo{pages}{765--773} (\bibinfo{year}{2005}).
\newblock \eprint{astro-ph/0411446}.

\bibitem{Lim12}
\bibinfo{author}{{Lim}, J.}, \bibinfo{author}{{Ohyama}, Y.},
  \bibinfo{author}{{Yan}, C.-H.}, \bibinfo{author}{{Dinh-V-Trung}} \&
  \bibinfo{author}{{Wang}, S.-Y.}
\newblock \bibinfo{title}{{A Molecular Hydrogen Nebula in the Central cD Galaxy
  of the Perseus Cluster}}.
\newblock \emph{\bibinfo{journal}{Astrophys. J.}} \textbf{\bibinfo{volume}{744}},
  \bibinfo{pages}{112} (\bibinfo{year}{2012}).

\bibitem{Salome06}
\bibinfo{author}{{Salom{\'e}}, P.} \emph{et~al.}
\newblock \bibinfo{title}{{Cold molecular gas in the Perseus cluster core.
  Association with X-ray cavity, H{\ensuremath{\alpha}} filaments and cooling
  flow}}.
\newblock \emph{\bibinfo{journal}{Astron. Astrophys.}} \textbf{\bibinfo{volume}{454}},
  \bibinfo{pages}{437--445} (\bibinfo{year}{2006}).
\newblock \eprint{astro-ph/0603350}.

\bibitem{Lim08}
\bibinfo{author}{{Lim}, J.}, \bibinfo{author}{{Ao}, Y.} \&
  \bibinfo{author}{{Dinh-V-Trung}}.
\newblock \bibinfo{title}{{Radially Inflowing Molecular Gas in NGC 1275
  Deposited by an X-Ray Cooling Flow in the Perseus Cluster}}.
\newblock \emph{\bibinfo{journal}{Astrophys. J.}} \textbf{\bibinfo{volume}{672}},
  \bibinfo{pages}{252--265} (\bibinfo{year}{2008}).
\newblock \eprint{astro-ph/0712.2979}.

\bibitem{Ho09}
\bibinfo{author}{{Ho}, I.~T.}, \bibinfo{author}{{Lim}, J.} \&
  \bibinfo{author}{{Dinh-V-Trung}}.
\newblock \bibinfo{title}{{Multiple Radial Cool Molecular Filaments in NGC
  1275}}.
\newblock \emph{\bibinfo{journal}{Astrophys. J.}} \textbf{\bibinfo{volume}{698}},
  \bibinfo{pages}{1191--1206} (\bibinfo{year}{2009}).
\newblock \eprint{astro-ph/0903.5411}.

\bibitem{Fabian11a}
\bibinfo{author}{{Fabian}, A.~C.} \emph{et~al.}
\newblock \bibinfo{title}{{The energy source of the filaments around the giant
  galaxy NGC 1275}}.
\newblock \emph{\bibinfo{journal}{Mon. Not. R. Astron. Soc.}} \textbf{\bibinfo{volume}{417}},
  \bibinfo{pages}{172--177} (\bibinfo{year}{2011}).
\newblock \eprint{astro-ph/1105.1735}.

\bibitem{Wilman05}
\bibinfo{author}{{Wilman}, R.~J.}, \bibinfo{author}{{Edge}, A.~C.} \&
  \bibinfo{author}{{Johnstone}, R.~M.}
\newblock \bibinfo{title}{{The nature of the molecular gas system in the core
  of NGC 1275}}.
\newblock \emph{\bibinfo{journal}{Mon. Not. R. Astron. Soc.}} \textbf{\bibinfo{volume}{359}},
  \bibinfo{pages}{755--764} (\bibinfo{year}{2005}).
\newblock \eprint{astro-ph/0502537}.

\bibitem{Scharwachter13}
\bibinfo{author}{{Scharw{\"a}chter}, J.}, \bibinfo{author}{{McGregor}, P.~J.},
  \bibinfo{author}{{Dopita}, M.~A.} \& \bibinfo{author}{{Beck}, T.~L.}
\newblock \bibinfo{title}{{Kinematics and excitation of the molecular hydrogen
  accretion disc in NGC 1275}}.
\newblock \emph{\bibinfo{journal}{Mon. Not. R. Astron. Soc.}} \textbf{\bibinfo{volume}{429}},
  \bibinfo{pages}{2315--2332} (\bibinfo{year}{2013}).
\newblock \eprint{astro-ph/1211.6750}.

\bibitem{Nagai19}
\bibinfo{author}{{Nagai}, H.} \emph{et~al.}
\newblock \bibinfo{title}{{The ALMA Discovery of the Rotating Disk and Fast
  Outflow of Cold Molecular Gas in NGC 1275}}.
\newblock \emph{\bibinfo{journal}{Astrophys. J.}} \textbf{\bibinfo{volume}{883}},
  \bibinfo{pages}{193} (\bibinfo{year}{2019}).
\newblock \eprint{astro-ph/1905.06017}.

\bibitem{Salome08}
\bibinfo{author}{{Salom{\'e}}, P.} \emph{et~al.}
\newblock \bibinfo{title}{{Observations of CO in the eastern filaments of NGC
  1275}}.
\newblock \emph{\bibinfo{journal}{Astron. Astrophys.}} \textbf{\bibinfo{volume}{483}},
  \bibinfo{pages}{793--799} (\bibinfo{year}{2008}).
\newblock \eprint{astro-ph/0804.0694}.

\bibitem{Salome11}
\bibinfo{author}{{Salom{\'e}}, P.} \emph{et~al.}
\newblock \bibinfo{title}{{A very extended molecular web around NGC 1275}}.
\newblock \emph{\bibinfo{journal}{Astron. Astrophys.}} \textbf{\bibinfo{volume}{531}},
  \bibinfo{pages}{A85} (\bibinfo{year}{2011}).
\newblock \eprint{astro-ph/1105.3108}.

\bibitem{Fabian17}
\bibinfo{author}{{Fabian}, A.~C.} \emph{et~al.}
\newblock \bibinfo{title}{{Do sound waves transport the AGN energy in the
  Perseus cluster?}}
\newblock \emph{\bibinfo{journal}{Mon. Not. R. Astron. Soc.}} \textbf{\bibinfo{volume}{464}},
  \bibinfo{pages}{L1--L5} (\bibinfo{year}{2017}).
\newblock \eprint{astro-ph/1608.07088}.

\bibitem{Churazov02}
\bibinfo{author}{{Churazov}, E.}, \bibinfo{author}{{Sunyaev}, R.},
  \bibinfo{author}{{Forman}, W.} \& \bibinfo{author}{{B{\"o}hringer}, H.}
\newblock \bibinfo{title}{{Cooling flows as a calorimeter of active galactic
  nucleus mechanical power}}.
\newblock \emph{\bibinfo{journal}{Mon. Not. R. Astron. Soc.}} \textbf{\bibinfo{volume}{332}},
  \bibinfo{pages}{729--734} (\bibinfo{year}{2002}).
\newblock \eprint{astro-ph/0201125}.

\bibitem{Hatch06}
\bibinfo{author}{{Hatch}, N.~A.}, \bibinfo{author}{{Crawford}, C.~S.},
  \bibinfo{author}{{Johnstone}, R.~M.} \& \bibinfo{author}{{Fabian}, A.~C.}
\newblock \bibinfo{title}{{On the origin and excitation of the extended nebula
  surrounding NGC1275}}.
\newblock \emph{\bibinfo{journal}{Mon. Not. R. Astron. Soc.}} \textbf{\bibinfo{volume}{367}},
  \bibinfo{pages}{433--448} (\bibinfo{year}{2006}).
\newblock \eprint{astro-ph/0512331}.

\bibitem{Qiu20}
\bibinfo{author}{{Qiu}, Y.}, \bibinfo{author}{{Bogdanovi{\'c}}, T.},
  \bibinfo{author}{{Li}, Y.}, \bibinfo{author}{{McDonald}, M.} \&
  \bibinfo{author}{{McNamara}, B.~R.}
\newblock \bibinfo{title}{{The formation of dusty cold gas filaments from
  galaxy cluster simulations}}.
\newblock \emph{\bibinfo{journal}{Nature Astronomy}}
  \textbf{\bibinfo{volume}{4}}, \bibinfo{pages}{900--906}
  (\bibinfo{year}{2020}).
\newblock \eprint{astro-ph/2005.00549}.

\bibitem{Gaspari17}
\bibinfo{author}{{Gaspari}, M.}, \bibinfo{author}{{Temi}, P.} \&
  \bibinfo{author}{{Brighenti}, F.}
\newblock \bibinfo{title}{{Raining on black holes and massive galaxies: the
  top-down multiphase condensation model}}.
\newblock \emph{\bibinfo{journal}{Mon. Not. R. Astron. Soc.}} \textbf{\bibinfo{volume}{466}},
  \bibinfo{pages}{677--704} (\bibinfo{year}{2017}).
\newblock \eprint{astro-ph/1608.08216}.

\bibitem{Bolatto13}
\bibinfo{author}{{Bolatto}, A.~D.}, \bibinfo{author}{{Wolfire}, M.} \&
  \bibinfo{author}{{Leroy}, A.~K.}
\newblock \bibinfo{title}{{The CO-to-H$_{2}$ Conversion Factor}}.
\newblock \emph{\bibinfo{journal}{Ann. Rev. Astron. Astrophys.}} \textbf{\bibinfo{volume}{51}},
  \bibinfo{pages}{207--268} (\bibinfo{year}{2013}).
\newblock \eprint{astro-ph/1301.3498}.

\bibitem{Bregman06}
\bibinfo{author}{{Bregman}, J.~N.}, \bibinfo{author}{{Fabian}, A.~C.},
  \bibinfo{author}{{Miller}, E.~D.} \& \bibinfo{author}{{Irwin}, J.~A.}
\newblock \bibinfo{title}{{On VI Observations of Galaxy Clusters: Evidence for
  Modest Cooling Flows}}.
\newblock \emph{\bibinfo{journal}{Astrophys. J.}} \textbf{\bibinfo{volume}{642}},
  \bibinfo{pages}{746--751} (\bibinfo{year}{2006}).
\newblock \eprint{astro-ph/0602323}.

\bibitem{Holtzman92}
\bibinfo{author}{{Holtzman}, J.~A.} \emph{et~al.}
\newblock \bibinfo{title}{{Planetary Camera Observations of NGC 1275: Discovery
  of a Central Population of Compact Massive Blue Star Clusters}}.
\newblock \emph{\bibinfo{journal}{Astron. J.}} \textbf{\bibinfo{volume}{103}},
  \bibinfo{pages}{691} (\bibinfo{year}{1992}).

\bibitem{Lim22}
\bibinfo{author}{{Lim}, J.}, \bibinfo{author}{{Wong}, E.},
  \bibinfo{author}{{Ohyama}, Y.} \& \bibinfo{author}{{Yeung}, M. C.~H.}
\newblock \bibinfo{title}{{Recent Formation of a Spiral Disk Hosting Progenitor
  Globular Clusters at the Center of the Perseus Brightest Cluster Galaxy. II.
  Progenitor Globular Clusters}}.
\newblock \emph{\bibinfo{journal}{Astrophys. J.}} \textbf{\bibinfo{volume}{927}},
  \bibinfo{pages}{138} (\bibinfo{year}{2022}).
\newblock \eprint{astro-ph/2203.04185}.

\bibitem{Nagai10}
\bibinfo{author}{{Nagai}, H.} \emph{et~al.}
\newblock \bibinfo{title}{{VLBI Monitoring of 3C 84 (NGC 1275) in Early Phase
  of the 2005 Outburst}}.
\newblock \emph{\bibinfo{journal}{Pub. Astron. Soc. Japan}} \textbf{\bibinfo{volume}{62}},
  \bibinfo{pages}{L11} (\bibinfo{year}{2010}).
\newblock \eprint{astro-ph/1001.3852}.

\bibitem{Kino21}
\bibinfo{author}{{Kino}, M.} \emph{et~al.}
\newblock \bibinfo{title}{{Morphological Transition of the Compact Radio Lobe
  in 3C 84 via the Strong Jet-Cloud Collision}}.
\newblock \emph{\bibinfo{journal}{Astrophys. J. Lett.}} \textbf{\bibinfo{volume}{920}},
  \bibinfo{pages}{L24} (\bibinfo{year}{2021}).

\bibitem{Hogan15}
\bibinfo{author}{{Hogan}, M.~T.} \emph{et~al.}
\newblock \bibinfo{title}{{A comprehensive study of the radio properties of
  brightest cluster galaxies}}.
\newblock \emph{\bibinfo{journal}{Mon. Not. R. Astron. Soc.}} \textbf{\bibinfo{volume}{453}},
  \bibinfo{pages}{1201--1222} (\bibinfo{year}{2015}).
\newblock \eprint{astro-ph/1507.03019}.

\bibitem{ODea84}
\bibinfo{author}{{O'Dea}, C.~P.}, \bibinfo{author}{{Dent}, W.~A.} \&
  \bibinfo{author}{{Balonek}, T.~J.}
\newblock \bibinfo{title}{{The 20 year spectral evolution of the radio nucleus
  of NGC 1275.}}
\newblock \emph{\bibinfo{journal}{Astrophys. J.}} \textbf{\bibinfo{volume}{278}},
  \bibinfo{pages}{89--95} (\bibinfo{year}{1984}).

\bibitem{Dutson14}
\bibinfo{author}{{Dutson}, K.~L.} \emph{et~al.}
\newblock \bibinfo{title}{{A non-thermal study of the brightest cluster galaxy
  NGC 1275 - the Gamma-Radio connection over four decades}}.
\newblock \emph{\bibinfo{journal}{Mon. Not. R. Astron. Soc.}} \textbf{\bibinfo{volume}{442}},
  \bibinfo{pages}{2048--2057} (\bibinfo{year}{2014}).
\newblock \eprint{astro-ph/1405.3647}.

\bibitem{Paraschos23}
\bibinfo{author}{{Paraschos}, G.~F.} \emph{et~al.}
\newblock \bibinfo{title}{{A multi-band study and exploration of the radio
  wave-{\ensuremath{\gamma}}-ray connection in 3C 84}}.
\newblock \emph{\bibinfo{journal}{Astron. Astrophys.}} \textbf{\bibinfo{volume}{669}},
  \bibinfo{pages}{A32} (\bibinfo{year}{2023}).
\newblock \eprint{astro-ph/2210.09795}.

\bibitem{Sault95}
\bibinfo{author}{{Sault}, R.~J.}, \bibinfo{author}{{Teuben}, P.~J.} \&
  \bibinfo{author}{{Wright}, M.~C.~H.}
\newblock \bibinfo{title}{{A Retrospective View of MIRIAD}}.
\newblock In \bibinfo{editor}{{Shaw}, R.~A.}, \bibinfo{editor}{{Payne}, H.~E.}
  \& \bibinfo{editor}{{Hayes}, J.~J.~E.} (eds.)
  \emph{\bibinfo{booktitle}{Astronomical Data Analysis Software and Systems
  IV}}, vol.~\bibinfo{volume}{77} of \emph{\bibinfo{series}{Astronomical
  Society of the Pacific Conference Series}}, \bibinfo{pages}{433}
  (\bibinfo{year}{1995}).
\newblock \eprint{astro-ph/0612759}.

\bibitem{Westmeier21}
\bibinfo{author}{{Westmeier}, T.} \emph{et~al.}
\newblock \bibinfo{title}{{SOFIA 2 - an automated, parallel H I source finding
  pipeline for the WALLABY survey}}.
\newblock \emph{\bibinfo{journal}{Mon. Not. R. Astron. Soc.}} \textbf{\bibinfo{volume}{506}},
  \bibinfo{pages}{3962--3976} (\bibinfo{year}{2021}).
\newblock \eprint{astro-ph/2106.15789}.

\bibitem{Lim17}
\bibinfo{author}{{Lim}, J.}, \bibinfo{author}{{Dinh-V-Trung}},
  \bibinfo{author}{{Vrtilek}, J.}, \bibinfo{author}{{David}, L.~P.} \&
  \bibinfo{author}{{Forman}, W.}
\newblock \bibinfo{title}{{The Role of Electron Excitation and Nature of
  Molecular Gas in Cluster Central Elliptical Galaxies}}.
\newblock \emph{\bibinfo{journal}{Astrophys. J.}} \textbf{\bibinfo{volume}{850}},
  \bibinfo{pages}{31} (\bibinfo{year}{2017}).
\newblock \eprint{astro-ph/1710.06186}.

\end{thebibliography}
\end{document}